\documentclass[twocolumn,showpacs,showkeywords,prb]{revtex4}
\usepackage{graphicx}
\usepackage{dcolumn}
\usepackage{amsmath}
\usepackage{latexsym}

\def\l{\left}
\def\r{\right}
\begin{document}

\title{On the self-interaction of dark energy in a ghost-condensate model}

\author{Gour Bhattacharya}
\email{drgour@yahoo.com}
\affiliation{Department of Physics, Presidency University, 86/1, College Street, Kolkata 700 073, India}
\altaffiliation{Visiting Associate, Inter University Centre for Astronomy and Astrophysics, Pune, India}
\author{Pradip Mukherjee}
\email{pradip@iucaa.ernet.in}
\affiliation{Department of Physics, Barasat Govt. College, 10 K. N. C. Road, Barasat, Kolkata 700 124, India}
\altaffiliation{Visiting Associate, Inter University Centre for Astronomy and Astrophysics, Pune, India}
\author{Anirban Saha}
\email{anirban@iucaa.ernet.in}
\affiliation{Department of Physics, West Bengal State University, Barasat, North 24 Paraganas, West Bengal, India}
\altaffiliation{Visiting Associate, Inter University Centre for Astronomy and Astrophysics, Pune, India}

\begin{abstract}
\noindent 
In a ghost-condensate model of dark energy the combined dynamics of the scalar field and gravitation is shown to impose non-trivial restriction on the self-interaction of the scalar field. Using this restriction we show that the choice of a zero self-interaction leads to a situation too restrictive for the general evolution of the universe. This restriction, obtained in the form of a quadratic equation of the scalar potential, is demonstrated to admit real solutions. Also, 
in the appropriate limit it reproduces the potential in the phantom cosmology.
\end{abstract}


\maketitle

\section{Introduction}
Recent cosmological observations 
indicate late-time acceleration of the observable universe \cite{NL1, NL2}. Why the evolution of the universe is interposed between an early inflationary phase and the late-time acceleration is a yet-unresolved problem. Various theoretical attempts have been undertaken to confront this observational fact. Although the simplest way to explain this behavior is the consideration of a cosmological constant \cite{wein}, the known fine-tuning problem \cite{DE} led to the dark energy paradigm. Here one introduces exotic dark energy component in the form of scalar fields such as quintessence \cite{quint1, quint2, quint3, quint4, quint5, quint6, quint7}, k-essence \cite{kessence1, kessence2, kessence3} etc. Quintessence is based on scalar field models using a canonical field with a slowly varying potential. On the other hand the models grouped under k-essence are characterized by noncanonical kinetic terms. 
A key feature of the k-essence models is that the cosmic acceleration is realized by the kinetic energy  of the scalar field. The popular models under this category include the phantom model, the ghost condensate model etc \cite{DE}. 

     It is well-known that the late time cosmic acceleration requires an exotic equation of state $\omega_{DE} < -\frac{1}{3}$. Current observations allow 
 $\omega_{DE} < -1$ which can be explained by considering  negative kinetic energy 
with a field potential.
The resulting phantom model \cite{phantom1, phantom2, phantom3, phantom4, phantom5, phantom6} is extensively used to confront cosmological observation \cite{phantom_obs1, phantom_obs2, phantom_obs3, phantom_obs4, phantom_obs5, phantom_obs6}. This model is however ridden with various instabilities as its energy density is unbounded. This instability can be eliminated in the so-called ghost-condensate models \cite{GC} by including a term quadratic in the kinetic energy. 
In this context let us note that to realize the late-time acceleration scenario some self-interaction must be present in the phantom model. In contrast, in the ghost-condensate models the inclusion of self-interaction of the scalar field is believed to be a matter of choice \cite{DE}.
This fact, though not unfamiliar, has not been emphasised much in the literature. 

Since very little is known about the nature of dark energy it may appear that the presence or otherwise of an interaction term in the ghost-condensate model may not be ascertained from any fundamental premise. However, in this letter we show that this issue can be settled by demanding a consistent scalar field dynamics. We establish here that this consistency requirement imposes non-trivial restriction on the choice of the self-interaction in the ghost-condensate model. Using this restriction we show that describing the general evolutionary scenario of the universe using a ghost-condensate without self-interaction may lead to too restrictive a situaton. Specifically, in the bouncing universe scenario \cite{ekpyrotic1, ekpyrotic2, ekpyrotic3, bounce1, bounce2, bounce3, bounce4} where the universe bounces from a contracting to an expanding phase, absence of self-interaction of the ghost-condensate is not admissible at all. Further, that a real solution for the self-interaction potential is compatible with the ghost field dynamics has been demonstraed using the restriction obtained here. It may also be noted that in the appropriate limit the ghost-condensate model is known to go over to the phantom model. Reassuringly, the restriction we have derived, reproduces the phantom potential \cite{gumjudpai} in the same limit. 

At this point it will be appropriate to describe the organisation of this letter. In section 2 the ghost condensate model is introduced where we include an arbitrary self-interaction potential. The equations of motion for the scalar field and the scale factor are derived. These equations exhibit the coupling between the scalar field dynamics and gravity. Expressions for the energy density and pressure of the dark energy components are computed. These expressions are used in section 3 to demonstrate that the requirement of consistency between the Friedman equations and the scalar field equations imposes nontrivial restriction on the self-interaction potential in the form of a quadratic equation. The consequences of this is discussed. The concluding remarks are contained in section 4. We use mostly positive signature of the metric.

\section{The ghost condensate model with self-interaction of the scalar field}
\label{model}
In this section we consider the ghost condensate model with a self-interaction potential $V(\phi)$.
The action is given by 
\begin{equation}
S=\int d^{4}x \sqrt{-g} \left[\frac{R}{2k^{2}}
+{\cal{L}}_{\phi}
+{\cal{L}}_{\rm{m}}\right], \label{ghost}
\end{equation}
where 
\begin{eqnarray}
{\cal{L}}_{\phi} &=& -X + \frac{X^{2}}{M^{4}} - V\left( \phi \right) \label{l}\\
X &=& -\frac{1}{2}g^{\mu \nu}\partial_{\mu}\phi \partial_{\nu}\phi
\label{kinetic}
\end{eqnarray}
$M$ is a mass parameter, $R$ the Ricci scalar and $G = k^{2}/8\pi$ the gravitational constant. The term ${\cal{L}}_{\rm{m}}$ accounts for the total (dark plus baryonic) matter content of the
universe, which is assumed to be a barotropic fluid with energy density $\rho_m$ and pressure $p_m$, and equation-of-state parameter $w_m=p_m/\rho_m$. We neglect the radiation sector for simplicity.

 The action given by equation (\ref{ghost}) describes a scalar field interacting with gravity. 
Invoking the cosmological principle one requires the metric to be 
of the Robertson-Walker (RW) form
\begin{equation}
ds^2=dt^2-a^2(t)\left[\frac{dr^2}{1-Kr^2}+r^2d\Omega_2^2\right],
\end{equation}
where $t$ is the cosmic time, $r$ is the spatial radial coordinate, $\Omega_2$ is the 2-dimensional unit sphere volume, $K$ characterizes the curvature of 3-dimensional space and
$a(t)$ is the scale factor.
The Einstein equations lead to the Friedmann equations 
\begin{eqnarray}
H^{2}&=&\frac{k^{2}}{3}\Big(\rho_{m}+\rho_{\phi}\Big)-
\frac{K}{a^2} \label{FR1}\\
\dot{H}&=&-\frac{k^{2}}{2} \Big(\rho_{m}+p_m+\rho_{\phi}+p_{\phi}\Big)+\frac{K}{a^2}, \label{FR2} 
\end{eqnarray}
In the above a dot denotes derivative with respect to $t$ and
$H\equiv\dot{a}/a$ is the Hubble parameter. In these expressions,
$\rho_{\phi}$ and $p_\phi$  are respectively the energy density
and pressure of the scalar field. The quantities 
$\rho_{\phi}$ and $p_\phi$ are defined through the symmetric energy-momentum tensor
\begin{equation}
T^{(\phi)}_{\mu\nu} = \frac{-2}{\sqrt{-g}}\frac{\delta}{\delta g^{\mu\nu}}\left({\sqrt{-g}} \right)
\end{equation} 
A straightforward calculation gives
\begin{equation}
T^{(\phi)}_{\mu\nu} = g_{\mu\nu}{\cal{L}}_{\phi} + \left(-1 + \frac{2X}{M^4}\right)\partial _\mu\phi\partial _\nu\phi
\end{equation}
Assuming a perfect fluid model we identify
\begin{eqnarray}
\rho_{\phi} &=& -X + \frac{3X^{2}}{M^{4}} + V\left( \phi \right) 
\label{density}\\
p_{\phi} &=& {\cal{L}}_{\phi} = -X + \frac{X^{2}}{M^{4}} - V\left( \phi \right) 
\label{pressure}
\end{eqnarray}
The equation of motion for the scalar field $\phi$ can be derived from the action (\ref{ghost}). Due to the isotropy of the FLRW universe the scalar field is a function of time only. Consequently, its equation of motion reduces to
\begin{equation}
\l(1 - \frac{3 \dot{\phi}^{2}}{M^{4}}\r)\ddot{\phi} + 3H\l(1 - \frac{\dot{\phi}^{2}}{M^{4}}\r)\dot{\phi}- \frac{dV}{d \phi} = 0. \label{eqm}
\end{equation}
As is well known the same equation of motion follows from the conservation of $T_{\mu\nu}$. Indeed under isotropy the equations (\ref{density}) and (\ref{pressure}) reduce to 
\begin{eqnarray}
\rho_{\phi} &=& - \frac{1}{2}\dot{\phi}^{2} + \frac{\dot{3\phi}^{4}}{4 M ^{4}} + V\left( \phi \right)
\label{rhophi}\\
p_{\phi} &=& - \frac{1}{2}\dot{\phi}^{2} + \frac{\dot{\phi}^{4}}{4 M ^{4}} - V\left( \phi \right)
\label{pphi}
\end{eqnarray}
From the conservation condition $\nabla_\mu T^{(\phi)\mu\nu} = 0$ we get 
\begin{equation}
\dot{\rho}_\phi+3H(\rho_\phi+p_\phi)=0, \label{rhodot}
\end{equation}
which, written equivalently in field terms gives equation (\ref{eqm}).

To complete the set of differential equations (\ref{FR1}), (\ref{FR2}), (\ref{rhodot}) we include the equation for the evolution of matter density
\begin{eqnarray}
\dot{\rho}_m+3H(1+w_m)\rho_m=0, \label{rhomdot}
\end{eqnarray}
where $w_{m}=p_{m}/\rho_{m}$ is the matter equation of state parameter. The solution to equation (\ref{rhomdot}) can immediately be written down as
\begin{equation}
\frac{\rho_m}{\rho_{m0}} = \l[\frac{a\l(t_{0}\r)}{a\l(t\r)}\r]^{n}, \label{rhom}
\end{equation}
where $n = 3 (1 + w_m) $ and $\rho_{m0} \geq 0$ is the value of matter density at present time $t_0$. Now, the set of equations (\ref{FR1}), (\ref{FR2}), (\ref{rhodot}) and (\ref{rhomdot}) must give the dynamics of the scalar field under gravity in a self-consistent manner. In the next section we demonstrate that this consistency requirement constrains the self-interaction $V\l(\phi\r)$ in (\ref{ghost}). 

\section{Restriction on the self-interaction of the scalar field}
We start by constructing two independent combinations of the pressure and energy density of the dark energy sector in terms of the Hubble parameter $H$, matter energy density $\rho_{m}$, matter equation of state parameter $w_{m}$ and curvature parameter $K$ using (\ref{FR1}), (\ref{FR2}) and (\ref{rhomdot})
\begin{eqnarray}
\rho_{\phi} + p_{\phi} =A &=& -\frac{2 \dot{H}}{k^{2}}-\frac{n}{3} \rho_{m}  + \frac{2K}{k^{2}a^{2}} \label{A} \\
\rho_{\phi} + 3p_{\phi} =B &=& -\frac{6 \ddot{a}}{k^{2}a} - \l(n-2\r) \rho_{m} \label{B} 
\end{eqnarray}
Using equations (\ref{rhophi}) and (\ref{pphi}), we rewrite these combinations in terms of the ghost condensate field derivative $\dot{\phi}$ and potential $V\left( \phi \right)$:
\begin{eqnarray}
\rho_{\phi} + p_{\phi} =A &=&  - \dot{\phi}^{2} + \frac{\dot{\phi}^{4}}{M ^{4}} \label{A1} \\
\rho_{\phi} + 3p_{\phi} =B &=&  - 2 \dot{\phi}^{2} + \frac{3\dot{\phi}^{4}}{2M ^{4}} - 2 V\left( \phi \right) \label{B1} 
\end{eqnarray}
Inverting these equations to write $\dot{\phi}^{2}$ and $\dot{\phi}^{4}$ in terms of $A$, $B$ and $V\left( \phi \right)$ and utilizing the algebraic identity $\l(\dot{\phi}^{2}\r)^{2} = \dot{\phi}^{4}$ we obtain the following quadratic equation 
\begin{eqnarray}
V^{2}\l(\phi\r) && + \l(B - \frac{3A}{2} + \frac{M^{4}}{4}\r) V\l(\phi\r) \nonumber \\
&& + \frac{\l(3A - 2B\r)^{2} - 4 M^{4}\l(A - B/2\r)}{16} = 0
\label{quad} 
\end{eqnarray}
This is the restriction on the choice of potential in the ghost-condensate model which has been indicated earlier. Note that the simple fact that $V\l(\phi\r)$ has to satisfy a restriction of this form implies that care must be taken in asserting the absence of the self-interaction term. We will presently discuss this and other issues related to equation (\ref{quad}) in the following. Meanwhile, observe that in the limit $M^{4} \to \infty$ the equation (\ref{quad}) reduces to 
\begin{eqnarray}
V\l(\phi\r) = A - B/2
\label{limit} 
\end{eqnarray}
Substituting for $A$ and $B$ in (\ref{limit}) and simplifying, we get 
\begin{eqnarray}
V\l(\phi\r) = \frac{1}{k^{2}}\l(3H^{2} + \dot{H} + \frac{2K}{a^{2}}\r) + \frac{n-6}{6} \rho_{m}
\label{limit1} 
\end{eqnarray}
where equations (\ref{A}) and  (\ref{B}) have been used. This reproduces the result for the potential in the phantom model \cite{gumjudpai}. 

Coming back to equation (\ref{quad}) let us first investigate the possibility of a vanishing self-interaction. Substituting $V\l(\phi\r) = 0$ we get 
\begin{eqnarray}
\frac{\l(3A - 2B\r)^{2}}{4 M^{4}} = \l(A - B/2\r)
\label{zero}
\end{eqnarray}
Since the left hand side is positive definite we immediately get the condition 
\begin{eqnarray}
\l(A - \frac{B}{2}\r) = \frac{1}{k^{2}}\l(3 H^{2} + \dot{H} + \frac{2K}{a^{2}}\r) + \frac{n-6}{6} \rho_{m} \ge 0 \nonumber \\
\label{zero1}
\end{eqnarray}
Assuming matter in the form of dust ($n = 3 $) in a universe with flat geometry ($K = 1$), this can be further simplified to 
\begin{eqnarray}
3 H^{2} + \dot{H} \ge \frac{k^{2}}{2} \rho_{m} > 0
\label{zero2}
\end{eqnarray}
Using $\dot{H} + H^{2} = \ddot{a}/a$ we reexpress this as
\begin{eqnarray}
 H^{2} >  - \frac{1}{2}  \frac{\ddot{a}}{a}
\label{zero3}
\end{eqnarray}
This condition appears to be too restrictive, in fact in the decelerating phase ($\ddot{a} < 0 $) this imposes a definite relation between $\dot{a}$ and $\ddot{a}$. Remembering that the Friedmann equation is of second order in time, there is no a priori reason that such constraint holds. Moreover, the ekpyrotic \cite{ekpyrotic1, ekpyrotic2, ekpyrotic3} and other bouncing theories \cite{bounce1, bounce2, bounce3, bounce4} of the early universe require that spacetime ``bounce'' from a contracting to an expanding phase, perhaps even oscillating cyclically [9, 10]. Clearly, during the switch over from expanding to contracting phase, $\dot{a} = 0$ but $\ddot{a} < 0$ and thus the condition (\ref{zero3}) is violated. 

The analysis detailed above demonstrates that in order to apply the ghost condensate model for general evolution of the universe a certain self-interaction should always be included. At this point one may wonder whethar the constraining equation (\ref{quad}) on $ V\l(\phi\r) $ at all allows a real solution. Solving (\ref{quad}) we get 
\begin{eqnarray}
V\l(\phi\r) = \l(\frac{3A - 2B}{4} -\frac{M^{4}}{8}\r) \pm \l\{\frac{M^{4}}{16}\l(\frac{M^{4}}{4} +A\r)\r\}^{\frac{1}{2}} 
\label{V} 
\end{eqnarray}
The reality condition is thus 
\begin{eqnarray}
\l(\frac{M^{4}}{4} +A \r) \ge 0     
\label{reality} 
\end{eqnarray}
That this condition is satisfied in general can be established explicitly if we substitute for $A$ from equation (\ref{A1}) which gives 
\begin{eqnarray}
\l(\frac{M^{4}}{4} +A \r)  = \frac{1}{M^{4}}\l(\dot{\phi}^{2} - \frac{M^{4}}{2}\r)^{2} \ge 0
\label{reality_check} 
\end{eqnarray}
This completes our argument in favour of including a self-interaction potential in ghost-condensate model.

\section{Conclusion}
We have considered the ghost-condensate model of dark energy with a self-interaction potential in a general FLRW universe with curvature $K$. The combined dynamics of dark energy and gravity leads to coupled differential equations involving the universal scale factor $a\l(t\r)$ and the scalar field $\phi$. The standard barotropic matter equation of state is assumed. Two independent combination of the pressure and energy density of the dark energy are expressed in terms of the observable quantities from the normal matter and gravity sector. These combinations are then used to impose a consistency condition which leads to a quadratic equation for the self-interaction $V\l(\phi\r)$. This equation is shown to admit real roots.
Also, in the appropriate limit it leads to the phantom model potential \cite{gumjudpai}. 

A very interesting consequence arises when we examine the plausibility of the choice of zero self-interaction. Using the quadratic equation satisfied by the self-interaction it has been demonstrated that this choice is too restrictive for the general evolution of the scale factor. In fact, the bouncing universe scenario disallows such a choice. Our analysis thus establishes that in the class of ghost-condensate models for general evolution of the universe a self-interaction of the dark energy must be included.

\section*{Acknowledgement} The authors would like to thank IUCAA, Pune, where part of the work was done.

\end{document}